\documentclass[12pt,letterpaper]{JHEP3}
\usepackage{amsmath,amssymb,amsfonts,xspace,mathrsfs}
\usepackage{epsfig}

\newcommand{\color}[6]{}  

\def\lfig#1#2#3#4#5{
 \begin{figure}
 \refstepcounter{figure}
 \label{#4}
 \addtocounter{figure}{-1}
 \epsfxsize=#3
 \centerline{\epsfbox{#2}}
 \vspace{#5}
 {\bf \caption{{\rm #1}}}
 \end{figure}
}

\def\Ref#1{(\ref{#1})}

\def\hb{\hbar}

\def\Im{\,{\rm Im}\,}
\def\Re{\,{\rm Re}\,}

\def\({\left(}
\def\){\right)}
\def\[{\left[}
\def\]{\right]}
\def\haf{\textstyle{1\over 2}}
\def\hf{\frac{1}{2}}

\newcommand{\de}{\mathrm{d}}

\newcommand{\I}{\mathrm{i}}

\newcommand{\p}{\partial}

\newcommand{\cQ}{\mathcal{Q}}
\newcommand{\cM}{\mathcal{M}}

\newcommand{\cX}{\mathcal{X}}

\newcommand{\cR}{\mathcal{R}}

\DeclareSymbolFont{AMSa}{U}{msa}{m}{n}
\DeclareSymbolFont{AMSb}{U}{msb}{m}{n}
\DeclareMathSymbol{\fieldR}{\mathalpha}{AMSb}{"52}

\newcommand{\cZ}{\mathcal{Z}}

\newcommand{\cU}{\mathcal{U}}

\newcommand{\pa}{\partial}

\newcommand{\IR}{\mathbb{R}}
\newcommand{\IC}{\mathbb{C}}

\newcommand{\txi}{{\tilde\xi}}

\newcommand{\CP}{\IC P^1}

\newcommand{\beq}{\begin{eqnarray}}
\newcommand{\eeq}{\end{eqnarray}}
\def\be{\begin{equation}}
\def\ee{\end{equation}}
\def\ba{\begin{align}}
\def\ea{\end{align}}
\def\bse{\begin{subequations}}
\def\ese{\end{subequations}}
\newcommand{\bea}[2]{\be\label{#2}\begin{array}{#1}}
\newcommand{\eea}{\end{array}\ee}

\def\ba{\bar a}

\def\bz{\bar z}

\def\bF{\bar F}
\def\bL{\bar L}
\def\bM{\bar M}

\def\o{\omega}
\def\ho{\hat\omega}

\def\txii#1{{\tilde\xi}^{[#1]}}
\def\talp{\tilde\alpha}
\def\ai#1{{\alpha}^{[#1]}}

\def\xii#1{\xi^{[#1]}}

\def\Hij#1{H^{[#1]}}

\def\tpm{t_{\pm 1}}
\def\tmp{t_{\mp 1}}

\def\xp{x_{_{+}}}
\def\xm{x_{_{-}}}
\def\xpm{x_{_{\pm}}}
\def\xmp{x_{_{\mp}}}

\def\XXint#1#2#3{{\setbox0=\hbox{$#1{#2#3}{\int}$}
\vcenter{\hbox{$#2#3$}}\kern-.5\wd0}}

\newcommand{\hU}{{\mathcal{U}}}

\def\cij#1{c}
\def\ci#1{c}

\def\ba{\bar a}

\def\bz{\bar z}
\def\bu{\bar u}

\def\varpi{t}

\def\varpiqk{\mathrm{t}}

\def\todaQ{T}

\def\qkm{\cQ}

\title{\bf\Large c-map as $c=1$ string}
\author{
Sergei Alexandrov
\\

{\it Universit\'e Montpellier 2 \& CNRS, Laboratoire Charles Coulomb UMR 5221, F-34095,
Montpellier, France}

\vspace*{2mm} {\tt e-mail: \email{Sergey.Alexandrov@univ-montp2.fr}} \vspace*{-3mm}

}

\abstract{We show the existence of a duality between the c-map space
describing the universal hypermultiplet at tree level
and the matrix model description of two-dimensional string theory
compactified at a self-dual radius and perturbed by a sine-Liouville potential.
It appears as a particular case of a general relation between the twistor description of
four-dimensional quaternionic geometries and the Lax formalism for Toda hierarchy.
Furthermore, we give an evidence that the instanton corrections to the c-map metric coming from NS5-branes
can be encoded into the Baker-Akhiezer function of the integrable hierarchy.}

\begin{document}

\section{Introduction}
\label{sec_intro}

Integrability is a very powerful tool to find exact solutions and it is always a great achievement
when a theory is proven to possess an integrable structure. A particularly interesting class of theories
where various appearances of integrability have been observed during the last few years is characterized
by the presence of $N=2$ supersymmetry. Remarkably, it appears both in the gauge
\cite{Donagi:1995cf,Nekrasov:2009rc,Gaiotto:2008cd,Gaiotto:2010be,Cecotti:2010fi}
and in the string theory context \cite{Aganagic:2003qj,Alexandrov:2010pp}.
However, an exhaustive explanation of these observations is missing so far.

In this note we reveal one more instance of an integrable structure in theories with $N=2$
supersymmetry. Namely, we observe that the four-dimensional quaternion-K\"ahler (QK) space
obtained by the local c-map construction \cite{Cecotti:1988qn,Ferrara:1989ik}
is in a certain sense dual to a sine-Liouville
perturbation of the $c=1$ string theory compactified at the self-dual radius.
The former represents the tree level moduli space of the so called universal hypermultiplet \cite{Strominger:1997eb}
arising in compactifications of type II strings on a rigid Calabi-Yau.
On the other hand, the latter is known to be an exactly solvable theory which is described by Toda integrable hierarchy
\cite{Dijkgraaf:1992hk,Alexandrov:2002fh}.
In particular, the lowest equation of this hierarchy is given by the Toda equation,
which is also known to parametrize four-dimensional quaternionic metrics with one isometry \cite{Przanowski:1991ru}.
The duality which we report here is based on the observation that the two systems
are described by the same solution of the Toda equation.

However, the duality goes much further as it allows to identify also
various quantities characterizing either the c-map or the $c=1$ string.
These identifications originate in a general correspondence
between a twistor description of quaternionic manifolds and the Lax formalism
for integrable hierarchies \cite{Guha:1997fz}.
The duality between the c-map and the $c=1$ string is just its simplest example.

The above mentioned correspondence is not at all surprising.
First of all, the existence of integrable hierarchies governing the four-dimensional hyperk\"ahler (HK) geometry
is very well known (see, for instance, \cite{Takasaki:1988cd,Dunajski:2000iq,dunajski2003tth}).
Moreover, recently it was realized that hyperk\"ahler manifolds with a rotational isometry
endowed with a hyperholomorphic line bundle can be associated with QK manifolds
of the same dimension also possessing an isometry \cite{Haydys,Alexandrov:2011ac}.
In particular, this QK/HK correspondence identifies the rigid and local c-map spaces
and makes a contact between QK geometry and integrability.
Although, this might seem to imply that the duality we propose is rather trivial from the mathematical point of view,
its physical realization is quite fascinating as it suggests that the hypermultiplet sector
of $N=2$ superstring compactifications and non-critical $c=1$ string theory are closely related to each other.
We are not aware of any examples where some perturbation of two-dimensional string theory
was related to the hypermultiplet moduli space.

Of course, this duality would become really interesting if it can be extended beyond the classical limit
and used to extract quantum corrections on one side from the known results on the other side.
We leave the systematic study of this problem for the future. Nevertheless, in this note we make the first
step in this direction. Namely, we observe that the holomorphic functions on the twistor space
of the c-map, which have been shown to generate NS5-brane instanton corrections to
the tree level universal hypermultiplet, can be interpreted in terms of the Baker-Akhiezer wave function
of the Toda hierarchy.
Together with previous findings \cite{Gaiotto:2008cd,Alexandrov:2010pp}, this suggests that
the whole non-perturbative picture describing $N=2$ string compactifications possesses an integrable structure
where the relations discussed here will find their natural place.

\section{Toda lattice hierarchy}
\label{TODA}

We start from a brief review of two-dimensional Toda lattice hierarchy \cite{Ueno:1984zz}
which provides the general framework for what we are going to discuss.
This integrable hierarchy describes many physical phenomena which, in particular, include
a certain class of deformations of $c=1$ string theory. More precisely,
it encompasses all perturbations generated by the spectrum of tachyon operators corresponding
to the $c=1$ string compactified on a circle \cite{Dijkgraaf:1992hk,Alexandrov:2002fh}.
Furthermore, the Lax formalism of this hierarchy turns out be closely related
to the matrix model description of the $c=1$ string, which is provided by Matrix Quantum Mechanics (MQM).
Therefore, first we present this formalism, then we explain its interpretation
from the MQM point of view, and finally we give a particular solution
of the Toda hierarchy which describes $c=1$ string theory with the so-called
sine-Liouville perturbation.
Although we do not need the Toda hierarchy in its full generality to explain the relation
with the c-map spaces, it is nevertheless useful to recall it as we want to discuss
further extensions and generalizations of the proposed duality.

\subsection{Lax formalism}
\label{laxform}

Let us first give a formal definition of the Toda lattice hierarchy.
To this end, introduce two semi-infinite series
\beq
L= r(s)\ho+\sum\limits_{k=0}^{\infty} u_k(s)\ho^{-k},
\qquad
\bL= \ho^{-1}r(s)+\sum\limits_{k=0}^{\infty} \ho^{k}\bu_{k}(s),
\label{Lax}
\eeq
where $s$ is a discrete variable labeling the nodes of an infinite
lattice, $\ho=e^{\hb\p/\p s}$ is the shift operator in $s$,
and the Planck constant $\hbar$ plays the role of a spacing parameter.
The operators \Ref{Lax} are called Lax operators.
Their coefficients $r$, $u_k$ and $\bu_{k}$ are taken to be functions
of two infinite sets of ``times'' $\{t_{\pm k}\}_{k=1}^{\infty}$.
To each time variable one associates a Hamiltonian $H_{\pm k}$
generating an evolution along $t_{\pm k}$
by the usual rule
\bea{rclcrcl}{evolH}
 \hb{\p L\over\p t_k} &=&[H_k,L], &\qquad&
 \hb{\p \bL\over \p t_k} &=&[H_k,\bL], \\
 \hb{\p L\over \p t_{-k}} &=&[H_{-k},L], &\qquad&
 \hb{\p \bL\over \p t_{-k}} &=&[H_{-k},\bL].
\eea
This system represents the Toda hierarchy if the Hamiltonians satisfy an additional requirement, namely
that they can be expressed through the Lax operators \Ref{Lax} as follows
\be
H_k=(L^k)_> +\hf(L^k)_0,\qquad
H_{-k}=(\bL^k)_< +\hf(\bL^k)_0,
\label{laxh}
\ee
where the symbol $(\ )_{\stackrel{>}{<}}$ means the positive (negative)
part of the series in the shift operator $\ho$ and $(\ )_0$ denotes
the constant part.
The integrability follows from the existence of an infinite set of commuting flows generated
by $\hb{\p \over \p t_k}-H_k$ whose commutativity is a consequence of \eqref{evolH}.

This definition shows that the Toda hierarchy is a collection of non-linear equations of
the finite-difference type in $s$ and differential with respect to $t_k$
for the coefficients $r(s,t)$, $u_k(s,t)$ and $\bu_k(s,t)$.
Its hierarchic structure is reflected in the fact
that one obtains a closed equation on the first coefficient $r(s,t)$
and its solution provides the input information for the following
equations. It is easy to show that $r(s,t)$ should satisfy a discrete version of the Toda equation
\be
\hb^2 {\p^2 \log r^{2}(s)\over \p t_1 \p t_{-1}}=2r^{2}(s)-
r^{2}(s+\hb)-r^{2}(s-\hb),
\label{todaeq}
\ee
which gives the name to the hierarchy.

For our purposes it is important to introduce
the following Orlov-Shulman operators
\cite{Orlov:1986xz}
\be
\label{ORSH}
M=\sum\limits_{k\ge 1}kt_k L^k+s+\sum\limits_{k\ge 1}v_k L^{-k},
\qquad
\bM=-\sum\limits_{k\ge 1}kt_{-k} \bL^k+s-\sum\limits_{k\ge 1}v_{-k} \bL^{-k}.
\ee
The coefficients $v_{\pm k}(s,t)$ are supposed to be found from the condition
on the commutators of $M$ and $\bM$ with the Lax operators
\be
[L,M]=\hb L,\qquad [\bL,\bM]=-\hb \bL.
\label{laxorsh}
\ee
The importance of the Orlov-Shulman operators comes from the fact that they can be considered
as perturbations of the simple operators of multiplication by
the discrete variable $s$, whereas the Lax operators appear to be a ``dressed" version of
the shift operator $\ho$. Indeed, if one requires that $v_{\pm k}$
vanish when all $t_{\pm k}=0$, then in this limit $M=\bM=s$, $L=\ho$ and $\bL=\ho^{-1}$.
Moreover, the commutation relations \Ref{laxorsh} can be recognized as
a dressed version of the trivial relation
\be
[\ho,s]=\hb\,\ho.
\label{osos}
\ee

Another application of the Orlov--Shulman operators is that they lead directly
to the notion of $\tau$-function.
Indeed, one can show that their coefficients must satisfy \cite{Takasaki:1994xh}
\be
{\p v_k\over \p t_l}={\p v_l\over \p t_k}.
\label{vttv}
\ee
This implies that there exists a generating function $\tau_s[t]$
of $v_{\pm k}$ such that
\be
v_k(s,t)=\hb^2\, {\p \log \tau_s[t]\over \p t_k}.
\label{tauvk}
\ee
It is called the {\it $\tau$-function} of Toda hierarchy.
It encodes all information about a particular solution.
For example, it allows to extract
the first coefficient in the expansion of the Lax operators
\be
r^2(s-\hb)={\tau_{s+\hb}\tau_{s-\hb}\over\tau_s^2}.
\label{taur}
\ee
Plugging this relation into \eqref{todaeq}, one obtains that the $\tau$-function
is subject to the Toda equation.
In physical applications the $\tau$-function usually coincides with
partition functions and the coefficients $v_{k}$
are the one-point correlators of the operators generating the
commuting flows $H_{k}$.

Finally, note that the Toda hierarchy can be equivalently represented as an eigenvalue problem
given by the following equations
\be
x\Psi=L\Psi,
\qquad
\hbar x\,\frac{\p\Psi}{\p x}=M\Psi,
\qquad
\hbar\,\frac{\p\Psi}{\p t_k}=H_k\Psi,
\ee
plus similar equations for $\bar\Psi$ where $L,M$ should be replaced by $\bL,-\bM$.
The eigenfunction $\Psi(x;s,t)$ is known as the {\it Baker-Akhiezer function}.
Often it has a direct physical interpretation and plays a very important role.

\subsection{String equations}

The Toda hierarchy provides a universal description for many integrable systems
because the equations of the hierarchy have many different solutions.
There are two ways to select a particular solution. The most obvious one is to provide
an initial condition which can be, for instance, the $\tau$-function
for vanishing times corresponding physically to the partition function
of a non-perturbed system. However, the Toda equations involve partial differential
equations of high orders and require to know not only the $\tau$-function at $t_{\pm k}=0$ but also
its derivatives. Therefore, it is not always clear whether the
non-perturbed function is sufficient to restore the full $\tau$-function.

Fortunately, there is another way to select a unique solution of
the Toda hierarchy. It relies on the use of the so-called {\it string equations}
which are some algebraic relations between the Lax and Orlov--Shulman operators.
The advantage of the string equations is that they represent, in a
sense, already a partially integrated version of the hierarchy
equations. For example, instead of solving a second order differential
equation on the $\tau$-function, they reduce the problem to certain algebraic and
first order differential equations.

It is important, however, that
the string equations cannot be chosen arbitrary because they should be consistent
with the commutation relations \eqref{laxorsh}. For example, if they are given by two
equations of the following type
\be
\bL=f(L,M),\qquad \bM=g(L,M),
\label{streq}
\ee
the operators defined by the functions $f$ and $g$ must satisfy
\be
[f(\ho,s),g(\ho,s)]=-\hb f.
\label{constreq}
\ee

\subsection{Dispersionless limit}
\label{dispers}

In this paper we are mostly interested only in the classical limit of the Toda hierarchy,
which is obtained by taking the spacing parameter $\hb$ to zero.
In this limit one recovers what is known as {\it dispersionless Toda hierarchy} \cite{Takasaki:1994xh}.
It is described by the same equations as above with the only difference
that all commutators should be replaced by Poisson brackets and
all operators are considered now as functions on the phase space
parametrized by $s$ and $\o$ with the symplectic structure
induced from \Ref{osos}
\be
\{\o,s\}=\o.
\label{oss}
\ee

As in the full theory, a solution of the dispersionless Toda hierarchy
is completely characterized by a dispersionless $\tau$-function.
In fact, it is better to talk about free energy since
it is the logarithm of the full $\tau$-function that
can be represented as a series in $\hb$
\be
\log\tau=\sum\limits_{n\ge 0}\hb^{-2+2n}F_n.
\label{sertauh}
\ee
The dispersionless limit is described by the first coefficient of this series,
the dispersionless free energy $F_0$. It satisfies the classical limit
of the Toda equation
\be
\p_{t_1}\p_{t_{-1}}F_0+e^{\p^2_s F_0}=0,
\label{hToda}
\ee
and can be selected from all solutions
by the same string equations \Ref{streq} as in the quantum case.

The quasiclassical asymptotics of the Baker-Akhiezer function is given by the usual WKB approximation.
Since $\Psi$ diagonalizes the Lax operator, it can be considered as a wave function representing the quantum state
of the system in the $L$-representation. The conjugate variable is provided by $M$ which implies that
\be
\Psi\sim \exp\left\{\frac{1}{\hbar}\int^{\log x} M\de \log L \right\},
\label{asymBA}
\ee
where $M$ is considered as a function of $L$, $s$ and $t_k$. Using \eqref{ORSH}, one finds
\be
\Psi=e^{\frac{1}{\hbar}\,S+O(\hbar^0)},
\qquad
S= \sum_{k\ge 1} t_k x^k+s\log x+\hf\,\phi-\sum_{k\ge 1}\frac{v_n}{n}\, x^{-n}.
\label{Sgenf}
\ee
The integration constant $\phi$ can be related to the free energy as $\phi(s,t)=-\p_s F_0$.

\subsection{Example: MQM with sine-Liouville perturbation}
\label{subsec-mqm}

Two-dimensional or non-critical $c=1$ string theory is known to be
an example of integrable model described by the Toda hierarchy. However,
this is difficult to see in its CFT formulation. On the other hand, the integrability
becomes transparent in its matrix model description in terms of Matrix Quantum Mechanics
(see \cite{Klebanov:1991qa,Alexandrov:2003ut} for a review).

To see how it appears, one should restrict to the singlet sector of MQM where
all angular degrees of freedom are integrated out. The remaining matrix
eigenvalues describe a system of free one-dimensional fermions in the inverse oscillator potential.
The target space description of string theory arises as an effective theory of collective
excitations of the free fermions. In particular, different backgrounds correspond to
different states of the Fermi sea: the simplest linear dilaton background is dual
to the ground state, whereas various perturbed backgrounds with a non-trivial
condensate of the tachyon field come from ceratin time-dependent states.

The most convenient way to describe the dynamics of these free fermions
is to use the light-cone coordinates \cite{Alexandrov:2002fh} $\xpm=\frac{x\pm p}{\sqrt{2}}$
where $(x,p)$ are the coordinates on the phase space of one fermion.
It is clear that $\xpm$ satisfy the canonical commutation relations
\be
\{ \xp,  \xm\}= {1}.
\label{cancom}
\ee
In the quasiclassical approximation the problem reduces to finding
the exact form of the profile of the Fermi sea in the phase space,
given its asymptotic form at $\xpm\to\infty$. Here we are interested in deformations
generated by the spectrum of the theory compactified on a circle of radius $R$.
Then the profile of the Fermi sea is determined by the compatibility of the following
two equations
\be
\xp\xm=M_\pm(\xpm)=
\pm {1\over R}\sum\limits_{k\ge 1}  k t_{\pm k} \  \xpm^{ k/R}  +\mu  \pm
{1\over R}\sum\limits_{k\ge 1} v_{\pm k}\  \xpm^{-k/R}.
\label{conteq}
\ee
Here $\xpm^{k/R}$ correspond to the tachyon vertex operators with momentum $k/R$,
$t_{\pm k}$ are their coupling constants, $\mu$ is the Fermi level, and $v_{\pm k}$
are to be found as functions of $t_{\pm k}$ and $\mu$.
For vanishing coupling constants, the equations \eqref{conteq} describe the hyperbola
$\xp\xm=\mu$ which is the trajectory of a free fermion in the inverse oscillator potential.

Comparing the r.h.s. of \eqref{conteq} with \eqref{ORSH}, one immediately sees that $M_\pm$
are related to the Orlov-Shulman operators. In fact, one can establish a precise correspondence
between all data of the Lax formalism and MQM quantities:
\begin{itemize}
\item
the lattice parameter $s$ is identified with the Fermi level $\mu$;
\item
the Lax operators coincide with the light-cone coordinates, $L=\xp^{1/R}$, $\bL=\xm^{1/R}$;
\item
the Orlov-Shulman operators encode the asymptotics of the Fermi sea and are given by $M=R M_+$, $\bM=R M_-$;
\item
the Baker-Akhiezer function coincides with the perturbed one-fermion wave function;
\item
the string equations are the same as the equations for the profile of the Fermi sea and
correspond to the choice $f=(M/R)^{1/R} L^{-1}$, $g=M$.
\end{itemize}
The commutation relation \eqref{cancom} then follows from a combination of \eqref{laxorsh}
with the string equations. Finally, the classical limit of the shift operator $\o$
turns out to be a uniformization parameter for the Riemann surface described by \eqref{conteq}.\footnote{In addition,
to describe the theory in Lorentzian signature, the Planck constant
should be continued analytically to imaginary values, $\hbar_{\rm Toda}=\I\hbar_{\rm MQM}$.\label{foot-MQMh}}

Note that the two Baker-Akhiezer functions, $\Psi$ and $\bar\Psi$, give one-fermion wave functions
in the two chiral representations, $\xp$ and $\xm$, respectively. The requirement that
they represent the same quantum state is precisely what generates the profile equations \eqref{conteq}.
Due to this condition, the two wave functions are related by a Fourier transform.

Restricting to a finite set of non-vanishing couplings $t_{\pm k}$, it is straightforward
to find an exact solution, i.e. an explicit form of the Lax and Orlov-Shulman operators
as well as the $\tau$-function of this system.
We are interested in the particular case $R=1$ and $t_{\pm k}=0$ for all $k\ge 2$.
The perturbation induced by the couplings $t_{\pm 1}$ is known to correspond to the sine-Liouville operator
on the string worldsheet, and the compactification radius $R=1$ is the self-dual point with respect to T-duality
which leads to an enhanced symmetry.
The corresponding solution of the Toda hierarchy is given by \cite{Alexandrov:2002fh}
\be
F_0=\frac12\,\mu^2\(\log\mu-\frac32\)-\mu t_1 t_{-1},
\qquad
\xpm =\sqrt{\mu}\,\o^{\pm 1}\mp \tmp.
\label{coefzom1}
\ee

\section{Four-dimensional QK spaces and integrability}

Now we turn to a seemingly completely different subject --- quaternionic geometry,
which is characterized by the existence of three (almost) complex structures $J^i$ satisfying the algebra of
quaternions and appears in two forms, hyperk\"ahler and quaternion-K\"ahler.
In fact, it is well known that in four-dimensions HK spaces possess an integrable structure,
which in the presence of an isometry is given by the dispersionless Toda hierarchy presented in the previous section.
Moreover, as we explained in the introduction, the QK/HK correspondence
allows to transfer these results to the realm of four-dimensional QK geometry.
In this section we establish precise relations between the quantities used to describe QK spaces
with one Killing vector and the ones entering the Lax formalism.

\subsection{Toda ansatz}

The most direct way to see the appearance of Toda in the description of QK spaces is to
recall that in four dimensions QK manifolds coincide with self-dual Einstein spaces with
a non-vanishing cosmological constant. Such spaces are known to be classified by
solutions of a single non-linear differential equation \cite{Przanowski:1984qq}.
In the presence of an isometry, it can be reduced to the three-dimensional continuous Toda equation
\be
\p_{z} \p_{\bar z}  \todaQ +\p_ \rho^2 \, {\rm e}^ \todaQ = 0,
\label{Toda}
\ee
which is identical to the equation \eqref{hToda} on the dispersionless free energy of Toda hierarchy.
In terms of the solution $\todaQ(\rho,z,\bar z)$, the metric is given by \cite{MR1423177}
\be
\label{dstoda}
\de s_\qkm^2 = -\frac{3}{\Lambda}\[
\frac{P}{\rho^2} \left( \de \rho^2 + 4 {\rm e}^ \todaQ \de z \de\bar z \right)
+ \frac{1}{P\rho^2}\,(\de \theta + \Theta )^2\] ,
\ee
where the isometry acts as a shift in the coordinate $\theta$.
Here $P\equiv 1- \haf\,\rho \pa_ \rho  \todaQ $,
$\Lambda$ is the cosmological constant, and $\Theta$ is a one-form such that
\be
\label{dth}
\de \Theta = \I (\pa_z P \de z - \pa_{\bar z} P \de \bar z)\wedge \de  \rho - 2\I\, \partial_\rho(P {\rm e}^ \todaQ)
\de z\wedge \de\bar z.
\ee
The integrability condition for \eqref{dth} follows from \eqref{Toda}.
Of course, not only the metric, but also all geometric data such as quaternionic two forms, Levi-Civita
connection, etc., can be expressed through $\todaQ$.
We refer to \cite{Alexandrov:2009vj} for their explicit expressions.

Comparing the Toda equations \eqref{Toda} and \eqref{hToda}, one immediately finds the following identifications
\be
z,\bz \leftrightarrow t_{\pm 1},
\qquad
\rho \leftrightarrow s,
\qquad
T \leftrightarrow \p_s^2 F_0.
\label{ident-T}
\ee
In principle, one can go further and find similar relations for other quantities.
However, as we already know, working with differential equations is not the best way to describe
a solution of the hierarchy. It is much better to use string equations which turn out to have
a certain twistorial interpretation \cite{Takasaki:1988cd,Guha:1997fz}.
Similarly, QK manifolds have a much more powerful description in terms of their twistor spaces \cite{Alexandrov:2008nk}.
As we will see now, the two become essentially identical.

\subsection{Twistor space and Lax formalism}
\label{sec_review}

The twistor space $\cZ$ of a QK manifold $\cM$ is a $\CP$ bundle over $\cM$,
whose connection is given by the $SU(2)$ part $\vec p$ of the Levi-Civita
connection\footnote{Recall that the holonomy group of a $4n$-dimensional QK manifold is contained in $Sp(n)\times SU(2)$.}
on $\cM$ and the $\CP$ fiber corresponds to the sphere of almost complex structures $J^i$.
The twistor space carries a so called {\it complex contact structure} \cite{MR664330,MR1327157},
which is represented by a holomorphic one-form $\cX$ such that the associated top-form $\cX\wedge (\de\cX)^n$
is nowhere vanishing. Locally it can be written as
\begin{equation}
\label{c-pot}
\cX^{[i]} = \frac{2}{\varpiqk}\, {\rm e}^{\Phi^{[i]}} \(\de \varpiqk + p^+ -\I p^3 \,\varpiqk + p^-\, \varpiqk^2\),
\end{equation}
where the index $\scriptstyle{[i]}$ labels open patches of a covering of $\cZ$,
$\varpiqk$ parametrizes the $\CP$ fiber, and the function $\Phi^{[i]}(x^\mu,\varpiqk)$
is known as the contact potential.
In the case we are interested in, when $\cM$ features one Killing vector, the contact potential is actually real, globally well defined,
and $\varpiqk$-independent so that one can put $\Phi^{[i]}=\phi(x^\mu)$.

Locally, it is always possible to choose such coordinates that the contact form $\cX$
acquires a standard form. Restricting to the case $\dim_\IR\cM=4$, this means that in a patch $\hU_i\subset\cM$
one can write
\begin{equation}\label{c-form}
\mathcal {X}^{[i]} = \de\ai{i} +\xii{i} \de\txii{i},
\end{equation}
where $(\xi,\txi,\alpha)$ are analogous to holomorphic Darboux coordinates of complex symplectic geometry.
Generically, these coordinates must be regular in $\hU_i$. The only exception is the coordinates defined around
the north ($\{\varpiqk=0\}\in\hU_+$) and the south ($\{\varpiqk=\infty\}\in\hU_-$) poles of $\CP$.
In these patches the Darboux coordinates are assumed to have the following expansions
\be
\begin{split}
\xii{\pm}(\varpiqk)=\cR\varpiqk^{\mp 1}&\,+\sum_{n=0}^{\infty}\xii{\pm}_n \varpiqk^{\pm n},
\qquad\quad
\txii{\pm}(\varpiqk)=\sum_{n=0}^{\infty}\txii{\pm}_n \varpiqk^{\pm n},
\\
&
\ai{\pm}(\varpiqk)=\sum_{n=0}^{\infty}\ai{\pm}_n \varpiqk^{\pm n}-2 c \log \varpiqk,
\end{split}
\label{expDqk}
\ee
where $\cR$ is a real function on $\cM$ and $c$ is a real number known as {\it anomalous dimension}.
Thus, $\xii{\pm}$ have a simple pole and $\ai{\pm}$ have a logarithmic singularity controlled by $c$.
The latter has an important physical meaning generating the one-loop correction to the local c-map \cite{Alexandrov:2008nk}.

On the overlap of two patches $\hU_i\cap\hU_j$, the two coordinate systems
are related by a complex contact transformation. Such transformations are generated by holomorphic functions
$\Hij{ij}(\xii{i},\txii{j},\ai{j})$, which we call transition functions.
Together with the anomalous dimension, they encode all geometric information about $\cM$ and its twistor space $\cZ$.
To extract the metric from them, the most non-trivial step is to find the Darboux coordinates as functions of $\varpiqk$
and coordinates on $\cM$.
Although this cannot be accomplished generically, it is possible to write certain integral equations
on the Darboux coordinates in terms of $\Hij{ij}$. We refrain from writing them explicitly and
refer to \cite{Alexandrov:2009zh,Alexandrov:2011va} for more details.

This twistor description has been related to the one based on the Toda ansatz \eqref{dstoda}
in \cite{Alexandrov:2009vj}.
For our purposes it is sufficient to provide the relation between the coordinates, contact and Toda potentials,
which read as follows
\be
\rho={\rm e}^\phi,
\qquad
z=\frac{1}{2}\,\txii{+}_0,
\qquad
\theta=\frac12\,\Im\ai{+}_0,
\qquad
\todaQ=2\log\cR.
\label{Backlund_tw}
\ee
Note that the only effect of the anomalous dimension is to shift the contact potential
introduced in \eqref{c-pot}, i.e. $e^\phi=e^\phi|_{c=0}-c$. The identification \eqref{Backlund_tw} nicely maps
this shift to the ambiguity in the solution of the Toda equation,
$\tilde \todaQ(\rho,z,\bz)=\todaQ(\rho+c,z,\bz)$.
Although the two solutions, $T$ and $\tilde T$, are related by a coordinate change,
due to the explicit dependence of the metric \eqref{dstoda} on $\rho$, the geometry of the corresponding QK manifold
is strongly affected by the parameter $c$.

After this very brief review of the twistor description of QK spaces,
one can easily see how it is embedded into the Lax formalism of the dispersionless Toda hierarchy.
Indeed, comparing the expansions \eqref{expDqk} with \eqref{Lax},
one observes a close similarity between the Lax operators and the Darboux coordinates
$\xii{\pm}$: they are both given by Laurent series which contain one singular term.
This similarity is further strengthen by the comparison of the coefficients of these singular terms.
Using \eqref{Backlund_tw} and \eqref{taur}, one finds
\be
\cR= e^{T/2}\leftrightarrow e^{\p_s^2 F_0/2}=r,
\ee
in agreement with \eqref{ident-T}.
Correspondingly, the phase space variable $\o$ should be identified with the $\CP$ coordinate,
$\o^{-1}\leftrightarrow\varpiqk$.
Besides, the symplectic structures derived from \eqref{laxorsh} and $\de\cX$ \eqref{c-form} suggest that
the Orlov--Shulman operators should be related to the Darboux coordinates $\txii{\pm}$.
Finally, it is well known (see, for instance, \cite{Takasaki:1988cd,Takasaki:1994xh})
that the string equations \eqref{streq} can be considered as gluing conditions
between the two patches around the north and south poles of $\CP$.
The condition \eqref{constreq} in turn requires that
the gluing conditions generate a symplectomorphism preserving the symplectic structure.
We summarize all the identifications in the following table:
\bea{ccc}{dict}
\omega^{-1} & \quad \longleftrightarrow\quad & \varpiqk
\\
\mbox{lattice variable } s & \quad \longleftrightarrow\quad & \rho=e^\phi
\\
\p_s^2 F_0 & \quad \longleftrightarrow\quad & T=2\log\cR
\\
\tpm   & \quad \longleftrightarrow\quad & z,\bz
\\
L, \bL   & \quad \longleftrightarrow\quad & \xii{\pm}
\\
M/L, \bM/\bL   & \quad \longleftrightarrow\quad & \txii{\pm}
\\
\mbox{string equations} & \quad \longleftrightarrow\quad & \mbox{gluing conditions}
\eea

These identifications are however a bit incomplete because we need also to match the subleading terms
in the expansions of the Lax operators and Darboux coordinates. This depends on the precise matching
of string equations with the gluing conditions and should be considered case by case.
In particular. in the next section we consider the situation of our special interest
where all identifications can be made very precise and explicit.

In addition, they say nothing about the role
on the Toda hierarchy side of the third Darboux coordinate $\alpha$.
The point is that, in the case with an isometry, this is a derived quantity,
namely, it can be given in terms of the transition functions and the explicit expressions for $\xi$ and $\txi$.
Therefore, it is not essential for defining the system. Nevertheless, we will see in section \ref{sec-NS5}
that it also has a counterpart in the Lax formalism being related to the quasiclassical asymptotics
of the Baker-Akhiezer function.

\section{c-map vs $c=1$ string}
\label{sect-UHM}

\subsection{Universal hypermultiplet at tree level}

Let us now consider a very special QK space obtained by the local c-map construction \cite{Cecotti:1988qn,Ferrara:1989ik},
which maps a special K\"ahler manifold characterized by a homogeneous holomorphic prepotential $F(X)$ to a QK manifold.
Physically, the local c-map produces the tree level metric on the hypermultiplet moduli space which describes
the low energy effective action of type IIA string theory compactified on a Calabi-Yau manifold $\mathfrak{Y}$.
Here we are interested in the case where the corresponding quaternionic manifold is four-dimensional.
In this case the local c-map gives the moduli space of one hypermultiplet known under the name of
universal hypermultiplet, which appears in the case where $\mathfrak{Y}$ is a rigid Calabi-Yau,
i.e. with vanishing Hodge number $h_{2,1}(\mathfrak{Y})$.

Since in our case the prepotential, we start with, is a holomorphic function of one variable and homogeneous of degree 2,
it is always quadratic, $F(X)=\frac{\tau}{4}\, X^2$. Although the modulus $\tau$ carries an important physical information
\cite{Bao:2009fg}, for simplicity we restrict to the case $\tau=-\frac{\I}{2}$.
Then the resulting quaternionic metric is given by \eqref{dstoda} where the potential $T$
should be chosen to be one of the most trivial solutions of the Toda equation \cite{Davidse:2005ef,Alexandrov:2006hx}
\be
T=\log \rho.
\label{cmapToda}
\ee
The coordinates $(\rho\!=\!e^{\phi},z,\theta)$ describe respectively the dilaton, a pair of RR-fields, and the NS-axion
dual in four dimensions to the antisymmetric $B$ field.

On the other hand, the twistor description of this space is based on the covering by two patches only:
$\cU_+$ covers the north hemisphere of $\CP$ and $\cU_-$ covers the south one.
The single transition function is determined by
the prepotential,\footnote{We adapted the normalizations to have simple duality relations.}
\be
\Hij{+-}=-4\Im F(\xi)=\xi^2.
\label{symp-cmap}
\ee
In this case the Darboux coordinates on the twistor space have been found in \cite{Alexandrov:2008nk} and,
being expressed in terms of coordinates appearing in the Toda ansatz \eqref{dstoda}, read as follows
\be
\begin{split}
& \qquad \xii{\pm}=-(z+\bz)+\sqrt{\rho}\(\varpiqk^{-1}-\varpiqk\),
\\
\txii{+} &\, =2\(z+\sqrt{\rho}\, \varpiqk\),
\qquad
\txii{-} =2\(-\bz+\sqrt{\rho}\,\varpiqk^{-1}\),
\\
\ai{\pm}=& \,2\I\theta\mp \(\rho-2(\Re z)^2\mp 4\sqrt{\rho}(\Re z) \varpiqk^{\pm 1} -\rho \varpiqk^{\pm 2}\).
\end{split}
\label{solDarb}
\ee

\subsection{Duality}

After all these preparations,
it is now trivial to establish the correspondence between the four-dimensional c-map and the perturbed $c=1$ string
at the self-dual radius whose solution is encoded in \eqref{coefzom1}.
Indeed, since the Fermi level $\mu$ is identified with the coordinate $\rho$,
both systems are described by the same solution of the Toda equation
\be
\p_\mu^2 F_0=T=\log\rho.
\ee
Moreover, it is trivial to check that the identifications \eqref{dict} imply
\be
\begin{split}
\xp=L& \,=\xii{+}+\hf\,\txii{+},
\qquad\
\xm=\bL=-\xii{-}+\hf\,\txii{-},
\\
M/L&\,=\frac{1}{2}\, \txii{+},
\qquad\qquad\qquad
\bM/\bL=\frac{1}{2}\, \txii{-}.
\end{split}
\label{relxLxi}
\ee
As it should be, the string equations \eqref{conteq} appear then as a consequence of the gluing condition
\be
\txii{+}-\txii{-}=-2\xii{\pm}.
\ee
These equations provide the completion of the identifications in the general dictionary \eqref{dict}
and establish the full equivalence of the two systems.

In fact, the relations \eqref{relxLxi} are not very illuminating. One can obtain much nicer identifications
if one introduces an additional patch on the twistor space of the c-map.
Let it be an intermediate patch $\cU_0$ around the equator of $\CP$ (see Fig. \ref{fig-sphere})
and the associated transition functions be \cite{Alexandrov:2008nk}
\be
\Hij{+0}=2\I F(\xi),
\qquad
\Hij{-0}=2\I \bF(\xi).
\label{symp-cmap2}
\ee
This patch becomes especially important when one incorporates instanton corrections
to the hypermultiplet moduli space \cite{Alexandrov:2008gh} and has an advantage
that the Darboux coordinates in $\hU_0$ transform nicely under
symplectic rotations \cite{Neitzke:2007ke}. They are given explicitly by
\be
\begin{split}
\xi \equiv\xii{0}&\, =-(z+\bz)+\sqrt{\rho}\(\varpiqk^{-1}-\varpiqk\),
\\
\txi\equiv \txii{0} &\, =(z-\bz)+\sqrt{\rho}\(\varpiqk^{-1}+\varpiqk\),
\\
\talp\equiv 2\I\(2\ai{0}+\xii{0}\txii{0}\) &\, =\sigma+4\I\sqrt{\rho}\(z\varpiqk^{-1}+\bz\varpiqk\),
\end{split}
\label{solDarbzero}
\ee
where we denoted $\sigma=-8\theta-2\I(z^2-\bz^2)$.
Applying \eqref{relxLxi}, one finds remarkably simple relations
between the Darboux coordinates $(\xi,\txi)$ and the coordinates on the MQM phase space $(x,p)$
\be
\xi=\xp-\xm=\sqrt{2}p,
\qquad
\txi=\xp+\xm=\sqrt{2}x.
\label{simplerel}
\ee

\lfig{Covering of $\CP$ by three patches, the associated transition functions,
and the two contours incorporating the effects of NS5-branes.}{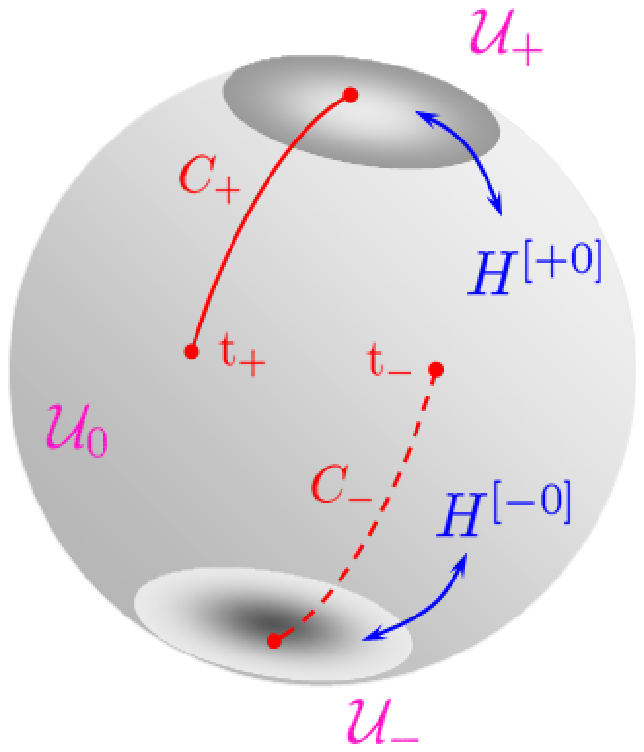}{8cm}{fig-sphere}{-1.2cm}

\section{NS5-branes and the Baker-Akhiezer function}
\label{sec-NS5}

In the previous section we clarified the meaning of the Darboux coordinates $\xi$ and $\txi$
under the correspondence with the $c=1$ string. Does the third coordinate $\alpha$ have some interpretation
on the dual side?
To answer this question, let us combine \eqref{c-form} and \eqref{c-pot} and integrate the resulting relation
over a contour on $\CP$ with a free upper limit. This leads to
\be
\alpha=2\rho\log\varpiqk+\int \xi\de \txi+{\rm const},
\ee
where the last term is independent of $\varpiqk$.
Then the dictionary \eqref{dict} implies that the second term should be related to
the phase of the quasiclassical Baker-Akhiezer function \eqref{asymBA}.

This expectation indeed turns out be correct and can be made quite precise in our particular case
characterized by the relations \eqref{simplerel}. Since the string equations imply that
$M/L=\xm$ and $\bM/\bL=\xp$, the phases of the two Baker-Akhiezer functions, $\Psi$ and $\bar\Psi$,
are given by\footnote{The constant term can be evaluated using \eqref{Sgenf}.}
\be
S(\xp)=\int\xm\de \xp,
\qquad
\bar S(\xm)=-\int\xp\de\xm.
\ee
Then using \eqref{coefzom1} and \eqref{solDarbzero}. it is easy to verify that
\be
\begin{split}
-\frac{\I}{8}\, \sigma+S(\xp)=&\, \frac{1}{8}\(-\I\talp+\txi^2-\xi^2\)-\rho\log\varpiqk,
\\
-\frac{\I}{8}\, \sigma+\bar S(\xm)=&\, -\frac{1}{8}\(\I\talp+\txi^2-\xi^2\)-\rho\log\varpiqk.
\end{split}
\label{phases-eval}
\ee
This confirms the identification of $\alpha$ (or a combination thereof with other coordinates)
with the WKB phase and shows that the coordinate $\sigma$ on the hypermultiplet moduli space
can be seen as a common constant imaginary phase in the Baker-Akhiezer functions.

Furthermore, the result \eqref{phases-eval} has a remarkable consequence.
It is based on the observation that the terms in the brackets on the r.h.s.
appear in the exponential of the holomorphic functions
\be
H_{k,\pm}^{\rm NS5} \sim
e^{\mp \pi \I k\tilde\alpha +\pi k\(\txi^2-\xi^2\)} ,
\label{NS5fun}
\ee
which have been shown to incorporate NS5-brane instanton
corrections to the classical geometry of the universal hypermultiplet \cite{Alexandrov:2009vj}.
Indeed, let $C_\pm$ be the two contours connecting $\varpiqk=0\, (\varpiqk=\infty)$
to the point $\varpiqk_{\pm}$, corresponding to the complex submanifold $\xi\mp\txi=0$ of the twistor space
and found to be
\be
\varpiqk_+=-\frac{z}{\sqrt{\rho}},
\qquad
\varpiqk_-=\frac{\sqrt{\rho}}{\bz}.
\ee
We depicted these contours in Fig. \ref{fig-sphere}. Then, being integrated over $C_\pm$,
the holomorphic functions $H_{k,\pm}^{\rm NS5}$ produce exponential corrections of the following form
\be
\int_{C_\pm} \frac{\de \varpiqk}{\varpiqk}\, H_{k,\pm}^{\rm NS5}
\sim e^{-4\pi k \(\rho+z\bz\)\mp \pi\I k\sigma}.
\label{integr}
\ee
This is a typical behavior of NS5-brane instantons since in our conventions $\rho\sim g_s^{-2}$ and
$\sigma$ is the axionic coupling associated with NS5-branes \cite{Becker:1995kb}.
The parameter $k$ is an integer interpreted as NS5-brane charge.

The holomorphic functions \eqref{NS5fun} are supposed to be considered as transition functions
modifying the complex contact structure of the twistor space of the c-map.\footnote{A new
feature here comparing to what we presented above is that these transition functions are not associated
to an overlap of any two open patches. In that case they would be integrated over a closed contour surrounding
one of the two open patches. Instead, the contours $C_\pm$ are open. Such contours typically arise
when the Darboux coordinates develop branch cut singularities. In particular, this is the case
for the twistor space description of D-brane instanton corrections to the local c-map \cite{Alexandrov:2008gh}
and BPS instanton corrections to the rigid c-map \cite{Gaiotto:2008cd}.}
Our results suggest that, up to a universal factor $\varpiqk^{\rho/\hbar}$, they can be identified with
(the quasiclassical limit of) the Backer-Akhiezer functions, provided
the Planck constant and the NS5-brane charge are related as
\be
\hbar^{-1}=8\pi k.
\label{identPlanck}
\ee
Amazingly, a similar relation between a quantization parameter and the NS5-brane charge
has been suggested in \cite{Alexandrov:2011ac} on the basis of a beautiful connection between
the geometric quantization of cluster varieties \cite{FGgeometricquantization},
motivic wall-crossing formula \cite{ks,Dimofte:2009tm}, and complex contact structure.

Let us recall also that in MQM the two Backer-Akhiezer functions can be seen as
perturbed one-fermion wave functions in the two chiral representations where $\xp$ and $\xm$
appear as diagonal operators, respectively.
Thus, the wave function in the right representation encodes the effects due to NS5-branes of charge $k$
and the (conjugate) wave function in the left representation does the same for anti-branes.
The precise relations read as follows
\be
H_{k,+}^{\rm NS5} \sim \varpiqk^{\rho/(\I\hbar_{\rm MQM})}\Psi(\xp)\, e^{-\pi \I k \sigma},
\qquad
H_{k,-}^{\rm NS5}\sim (\varpiqk^{\rho/(\I\hbar_{\rm MQM})}\bar\Psi(\xm))^* e^{\pi \I k \sigma},
\label{holPsirel}
\ee
where the complex conjugation does not act on the fields and at the end one should take $\hbar_{\rm MQM}=(8\pi \I k)^{-1}$
(see footnote \ref{foot-MQMh}). This identification is in the beautiful agreement with
a general expectation that the NS5-brane partition function should behave as a wave function
\cite{Witten:1997hc,Dijkgraaf:2002ac,Pioline:2009qt,Alexandrov:2010ca}
and, furthermore, suggests that NS5-branes might have a fermionic nature.

Moreover, the duality with MQM provides an interesting interpretation for the contours $C_\pm$.
In \cite{Alexandrov:2009vj} they have been chosen by hand to produce the correct instanton effects \eqref{integr}
and their meaning was completely unclear.
On the other hand, in terms of the MQM chiral coordinates, these contours turn out to be very simple: they join $\xpm=\infty$
to $\xmp=0$. The starting points correspond to the asymptotics of the Fermi sea, whereas the end points
are nothing else but the intersection points of the Fermi level with the lines of the perturbative stability.
Namely, the inverse oscillator potential $V=-\hf\, x^2$ has two infinite ``wells". Ignoring non-perturbative effects,
the movement of a particle is confined to one of them provided its initial conditions satisfy $x^2-p^2=2\xp\xm>0$.
Thus, the contours $C_\pm$ correspond to the parts of the Fermi profile which bounds the fermions
leaking to the other side of the potential. We demonstrated this situation on Fig. \ref{fig-sea}.

\lfig{The movement of free fermions in the inverse oscillator potential. The second picture shows the profile
of the Fermi sea on the two sides of the potential for the choice of the couplings $t_{\pm 1}=\mp 1$.
$\varpiqk_\pm$ are the points of intersection with the lines $\xmp=0$. They bound the parts of the Fermi sea
(marked by a darker color),
with the phase space coordinates satisfying $\xp\xm<0$, which leave to (come from) the opposite side of the potential.
This is illustrated on the third picture which shows the time evolution of the Fermi sea from the second picture.
One can see how a part of the sea leaks from the right to the left.}{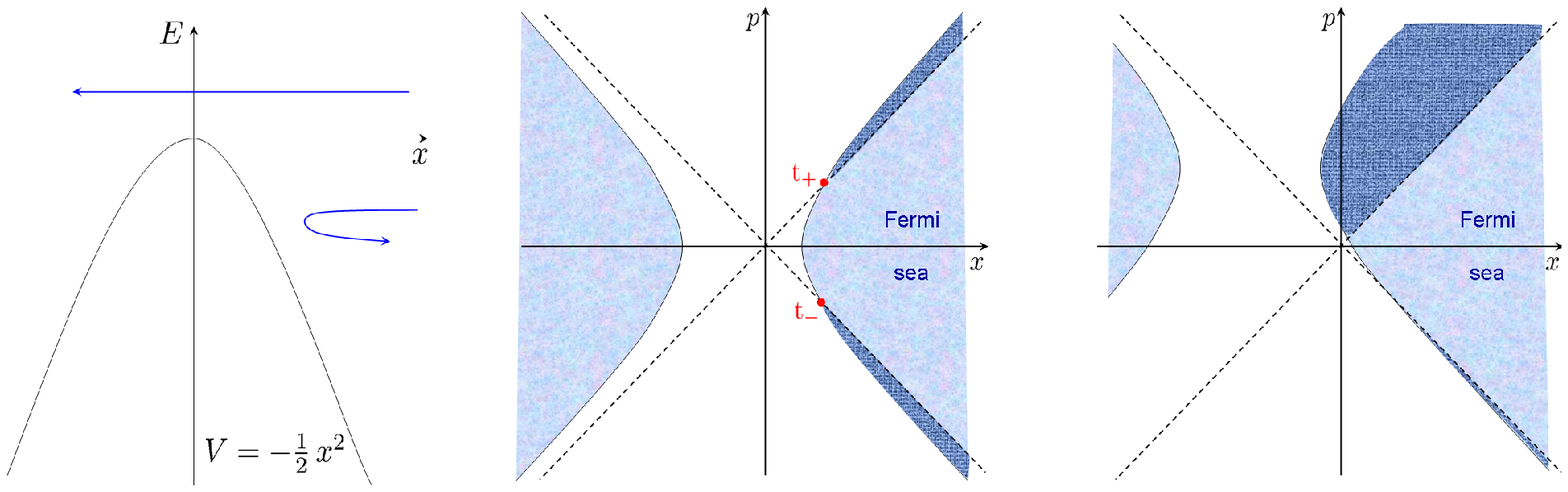}{17.6cm}{fig-sea}{-0.8cm}

\subsection{Inclusion of the one-loop correction}

It is very easy to include in the above description the one-loop $g_s$-correction to the universal hypermultiplet
geometry \cite{Antoniadis:2003sw,Anguelova:2004sj}. As was shown in \cite{Alexandrov:2008nk},
in the twistor formalism it appears as a non-vanishing anomalous dimension $c$,
which is fixed by the Euler characteristic of the Calabi-Yau, $c=-\frac{\chi_\mathfrak{Y}}{192\pi}$.
This anomalous dimension has only two effects: it adds a logarithmic term to $\alpha$ (see \eqref{expDqk})
and shifts the contact potential by a constant.
As discussed below \eqref{Backlund_tw}, this amounts to shifting in all above expressions $\rho$ by the constant $c$.
In particular, the one-loop corrected metric corresponds to the solution of the Toda equation $T=\log(\rho+c)$
and therefore leads to a modification of our dictionary by setting $\mu=\rho+c$.
The only place where $\rho$ remains unshifted is the relation \eqref{phases-eval} between $\alpha$
and the WKB phases since the shift is canceled by the logarithmic correction to $\alpha$.

The relation between the NS5-brane instantons and the Baker-Akhiezer function works in a similar way as above.
One should only take into account that the function \eqref{NS5fun}
should be multiplied by $(\txi\mp\xi)^{8\pi c k}$ \cite{Alexandrov:2009vj}.
As a result. the relation \eqref{holPsirel} generalizes to
\be
H_{k,+}^{\rm NS5} \sim (\varpiqk^{\rho}\xm^c)^{8\pi k}\Psi(\xp)\, e^{-\pi \I k \sigma}.
\ee

\section{Discussion}

In this note we proposed a correspondence between the QK space given by the local c-map
and the matrix model description of the $c=1$ string theory
with the sine-Liouville perturbation compactified at the self-dual radius.
This duality is based on a dictionary between the twistorial description
of four-dimensional quaternionic spaces having an isometry
and the Lax formulation of the dispersionless Toda hierarchy.
Under this duality, the coordinates parametrizing the quaternionic space appear as couplings of
the $c=1$ string theory, whereas the metric is encoded in the $\tau$-function of the hierarchy.
Besides, the coordinate parametrizing the $\CP$ fiber of the twistor space
appears on the matrix model side as a uniformization parameter of the Riemann surface
given by a complexification of the profile of the Fermi sea build from the free fermions of the matrix model.

While this curious duality is very interesting, it would become useful if it
can be applied to extract some information about physical effects on one side from
the known results on the other side. In particular, it is very important to understand whether
it can be extended to include quantum effects.
While the inclusion of the one-loop correction to the c-map is almost trivial, we also
provided an identification of the one-instanton corrections due to NS5-branes with
the Baker-Akhiezer function of the Toda hierarchy, which is realized in the $c=1$ string theory
as the wave function of a free fermion in the chiral representation.
This relation points towards a fermionic interpretation of NS5-branes which might have important implications
for the study of their dynamics. In particular, one might expect that multi-instanton effects
should be expressible in terms of a Slater determinant.

On the other hand, it is not clear how D-brane instantons to the hypermultiplet geometry
can fit into this correspondence (see \cite{Alexandrov:2011va} for a review
of the current understanding of the non-perturbative effects in the hypermultiplet sector).
The problem is that there are two string couplings in the game which have different scaling.
One is the coupling constant of the $c=1$ string theory which scales as $g_s^{c=1}\sim \mu^{-1}$.
The second string coupling describes quantum corrections to the hypermultiplet moduli space and
is set by the dilaton to $g_s\sim\rho^{-1/2}$. Recalling that $\mu=\rho$, one obtains that,
if D-instanton corrections to the c-map have a counterpart,
it should scale as $e^{-1/\sqrt{g_s^{c=1}}}$, which seems to be rather strange.

Moreover, one of the key ingredients determining the structure of instanton corrections
on the hypermultiplet moduli space is the Heisenberg symmetry. In particular, it requires that the coordinate $z$,
describing the RR-fields, lives on a torus.
Our correspondence then implies that the sine-Liouville couplings $t_{\pm 1}$ should be periodic.
It is not clear at all how such periodicity condition can arise in the $c=1$ string theory.

Thus, if the duality can be extended to include quantum effects, it should involve
a non-trivial deformation of the $c=1$ string and may even require to replace it by a different physical system.
Some indications towards this conclusion can be seen already here.
First of all, note that the NS5-brane instantons \eqref{integr} have the same strength as the tunneling effects in
the free fermionic system given by MQM. Due to this, one may expect that the fermions should fill
both ``wells" of the potential,
which leads to consideration of the supersymmetric 0B two-dimensional string theory
\cite{Douglas:2003up}. Second, there is a problem that the $c=1$ string and the c-map impose different reality conditions
on the Toda times: $t_{\pm 1}^*=t_{\pm 1}$ and $t_{1}^*=t_{-1}$, respectively.
This hints that the proper system dual to the c-map is not the $c=1$ string, but
the so-called Normal Matrix Model. It is given by the same solution of the Toda hierarchy, but describes
a different real section of the same holomorphic structure \cite{Kazakov:2002yh,Alexandrov:2003qk}.

It is important also to understand whether the presented results can be extended beyond the restriction to four dimensions.
The main complication appearing in higher dimensions is that
quaternionic geometries do not coincide anymore with self-dual Einstein spaces.
Nevertheless, one may hope that the Toda hierarchy still describes them upon switching on higher times
$t_{\pm k}$. Even if this is not the case, there is still a possibility that
an extension of this hierarchy to a more general one will do the job.

So far we were looking for an interpretation of the physical effects in the hypermultiplet sector of type II string theory
from the point of view of the $c=1$ string or its MQM realization.
One can consider also the opposite problem.
In particular, in the $c=1$ string theory with the sine-Liouville perturbation the
non-perturbative corrections have been studied in
\cite{Alexandrov:2003nn,Alexandrov:2004cg}.\footnote{It should be noted that most of the studies of $c=1$ string
theory with the sine-Liouville perturbation have been done for arbitrary compactification radius $R$.
At the same time, the limit $R\to 1$ is singular and a special care is required to extract finite results
both in the CFT and MQM approaches \cite{Alexandrov:2005gf}.}
One of the nice aspects of the resulting picture is that the instantons
are associated with singularities of the Riemann surface already mentioned above \cite{Seiberg:2003nm,Alexandrov:2004ks}.
It would be interesting to see whether such geometric interpretation can be translated to
the twistor space of the non-perturbative moduli spaces in $N=2$ theories.

Another issue to be understood is that on the Toda or $c=1$ string side, the presented duality involves so far only
the dispersionless limit. As was reviewed in section \ref{laxform}, there is a natural quantization
of this classical structure generating a perturbative expansion in $\hb$. In $c=1$ string
theory the $\hb$-corrections are interpreted as contributions from worldsheets of higher genera.
A natural question is: Is there a meaning of a similar deformation for quaternionic geometries?

There are several indications that this deformation corresponds to the inclusion of NS5-brane instantons.
Indeed, the Heisenberg symmetry mentioned above, in the presence of a non-vanishing NS5-brane charge,
becomes equivalent to the quantum torus algebra \cite{Alexandrov:2010ca}.
Therefore, one can expect that switching on this charge corresponds
to promoting certain structures from classical to quantum and
this quantization is equivalent to the one in the Toda hierarchy.
This expectation is strengthened by the observation that our duality implies the relation \eqref{identPlanck}
between the NS5-brane charge and the Planck constant.
Note also that the Toda hierarchy has been shown to incorporate the quantum torus algebra in several
physically relevant situations \cite{Takasaki:2011yz}.

\section*{Acknowledgements}

The author is grateful to Daniel Persson, Boris Pioline, Philippe Roche, Henning Samtleben, and especially to Stefan Vandoren
for many useful discussions.


\providecommand{\href}[2]{#2}\begingroup\raggedright\endgroup

\end{document}